\documentclass[11pt,twoside]{article}
\usepackage{amsmath} 
\usepackage{booktabs}
\usepackage{asp2010}

\resetcounters

\begin{document}

\title{Combining and comparing astrometric data from different epochs: A case study with Hipparcos and Nano-JASMINE}
\author{Daniel~Michalik$^1$, Lennart~Lindegren$^1$, David~Hobbs$^1$, Uwe~Lammers$^2$, and Yoshiyuki~Yamada$^3$}
\affil{$^1$Lund Observatory, Lund University, Box 43, SE-22100 Lund, Sweden\\e-mail: {\tt daniel.michalik, lennart, david@astro.lu.se}}
\affil{$^2$European Space Agency (ESA/ESAC), P.O. Box 78, ES-28691 Villanueva de la Ca\~{n}ada, Madrid, Spain, e-mail: {\tt uwe.lammers@sciops.esa.int}}
\affil{$^3$Department of Physics, Kyoto University, Oiwake-cho Kita-Shirakaw Sakyo-ku, Kyoto, 606-8502 Japan, e-mail: {\tt yamada@amesh.org}}

\begin{abstract}
The Hipparcos mission (1989-1993) resulted in the first space-based
stellar catalogue including measurements of positions, parallaxes and
annual proper motions accurate to about one milli-arcsecond. More space
astrometry missions will follow in the near future. The ultra-small
Japanese mission Nano-JASMINE (launch in late 2013) will determine positions
and annual proper motions with some milli-arcsecond accuracy. In
mid 2013 the next-generation ESA mission Gaia will deliver some tens of
micro-arcsecond accurate astrometric parameters. Until the final Gaia catalogue is published in early 2020 the best way
of improving proper motion values is the combination of positions from
different missions separated by long time intervals. Rather than
comparing positions from separately reduced catalogues, we propose an optimal method to combine the information from the different data
sets by making a joint astrometric solution. This allows to obtain good results even when each data set
alone is insufficient for an accurate reduction. 
We demonstrate our method by combining Hipparcos and simulated Nano-JASMINE data in a joint solution. 
We show a significant improvement over the conventional catalogue combination.
\end{abstract}

\section{Introduction \label{sect:intro}}
Stellar proper motions have traditionally been computed by comparing positional
catalogues based on observations made at different epochs (typically separated
by several decades). Parallaxes were either ignored in this
process, or determined by quite different instruments and methods.                                                                                                                                                                   
With the advent of Hipparcos and space astrometry, it has become necessary to
treat the determination of positions, proper motions, and parallaxes in a
unified manner, i.e., in a single least-squares solution. This was the
principle used for the construction of the Hipparcos and Tycho catalogues
\citep{1997ESASP1200.....P}, as well as for the new reduction of the Hipparcos
data \citep{2007ASSL..350.....V}, and it will be used for the Gaia mission and
other planned space astrometry projects.                                                                                                                                                                                                            
In the future we will therefore have access to several independent astrometric
catalogues, one for each space project. Improved proper motions can again be
computed by comparing the positions in catalogues at different epochs.
However rather then combining the results of the catalogues in the conventional manner, we propose to make
a {\em joint solution} of the data from the missions. In this paper we explore possible
advantages of this approach, combining the Hipparcos Catalogue
and simulated data from the Japanese Nano-JASMINE mission as a study case.     

Nano-JASMINE (launch planned for late 2013) is an ultra-small Japanese satellite to measure the astrometric parameters of about one million celestial
sources to 12th magnitude. The expected accuracy of the positions, parallaxes
and annual proper motions of magnitude 7.5 objects is about $3$ mas.
Nano-JASMINE is a very small technological demonstrator for bigger follow-up
missions.  Its data will be reduced using the Astrometric Global Iterative
Solution (AGIS) being developed for the Gaia mission at ESA/ESAC and Lund
Observatory \citep{2011A&A_AGIS}.

\section{Theory}
\label{sec:theory1}
The estimation of stellar astrometric parameters from observational data
can be cast as a linear least-squares problem. 
The optimum estimate of the unknowns $\boldsymbol{x}$ is obtained by solving
the normal equations $\boldsymbol{N}\boldsymbol{x} = \boldsymbol{b}$ where the covariance of $\boldsymbol{x}$ is given by
$\boldsymbol{N}^{-1}$ and where $\boldsymbol{b}$ are the residuals.
In the conventional approach of combining two astrometric catalogues the
least-squares solution of both data sets is done independently and the
combination is done a posteriori. 
If $\sigma_1$ and $\sigma_2$ are the accuracies in the two catalogues,
   positions and parallaxes improve as $\sigma^{-2} = \sigma_1^{-2} +
   \sigma_2^{-2}$ by computing weighted means of the values of the two
   catalogues. Proper motions improve as $\sigma_{pm} = (\sigma_{pos1}^2 +
	   \sigma_{pos2}^2)^{1/2} / \Delta T$ by taking the position difference divided by
	   the epoch difference $\Delta T$.
Instead we propose to combine the normal equations of both missions {\it before} solving
\begin{equation}
(\boldsymbol{N_1} + \boldsymbol{N_2}) \boldsymbol{x} = \boldsymbol{b_1} + \boldsymbol{b_2} ~~~~~ \rightarrow \boldsymbol{Nx} = \boldsymbol{b},
\end{equation}
allowing us to retrieve directly $\boldsymbol{\hat{x}}_{joint}$ of the combined catalogues.
This can also be understood in terms of Bayesian estimation (assuming multivariate Gaussian parameter errors), 
with $\boldsymbol{N}_1, \boldsymbol{b}_{1}$ presenting the prior information, $\boldsymbol{N}_2, \boldsymbol{b}_{2}$ the new data, and $\boldsymbol{N},\boldsymbol{b}$ the posterior information. Even if each group is not solvable on its own, the combined data may be solvable. 

\section{Simulations}
Simulations are carried out using AGISLab, a software package aiding the
development of algorithms for the data reduction of Gaia, developed at Lund
Observatory. For simulations the Hipparcos catalogue is used to generate observations. Three catalogues are created (see Figure \ref{michalik_fig:catalogues}):
\begin{figure}
\begin{center}
\includegraphics[width=0.8\textwidth]{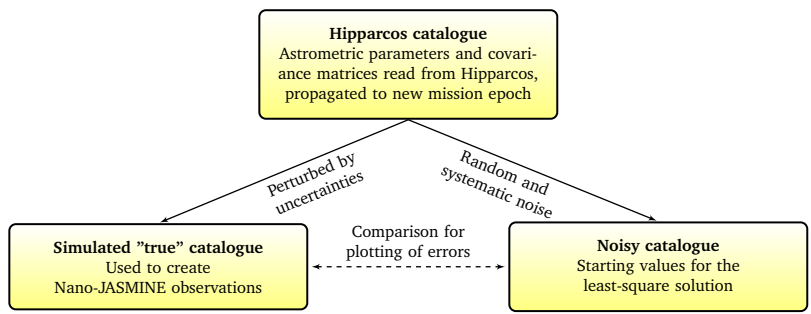}
\caption{Relationships between catalogues}\label{michalik_fig:catalogues}
\end{center}
\end{figure}

{\em The Hipparcos catalogue} contains the source parameters read from the
Hipparcos catalogue file including their covariance matrix.
The Hipparcos data are available for the reference epoch J1991.25. To combine them
with simulated Nano-JASMINE data the source parameters and covariance
matrix have to be propagated to the mid mission reference epoch (which for Nano-JASMINE
is expected to be J2015). 

{\em The noisy catalogue} contains the starting values for the data processing.
It is created by perturbing the original sources with random noise of a given
amplitude.

{\em The simulated ``true" catalogue} defines the sources used for creating the simulated true observations
(which may be perturbed subsequently). For the real mission the true catalogue
is not known. The Hipparcos sources give the observed astrometric parameters and
their uncertainties. The true parameters of the sources are expected to be at
an unknown position within the error space of the original sources. 
The five astrometric parameters of the simulated true catalogue require five independent Gaussian variates scaled by the square
root of the covariance matrix $\boldsymbol{C}$. 
To find a unique square root we use the lower triangular matrix $\boldsymbol{L}$ resulting from the
Cholesky decomposition $\boldsymbol{C} =  \boldsymbol{L}\boldsymbol{L}^T$. Then $\boldsymbol{e} = \boldsymbol{L} \boldsymbol{g}$, where $\boldsymbol{g}$ is a vector of five independent unit Gaussian values and $\boldsymbol{e}$ is the resulting vector of errors to be applied to the astrometric parameters.  Since $E(\boldsymbol{g}\boldsymbol{g}^T) = \boldsymbol{I}$, where $E(\ldots)$ denotes the expectation value and $\boldsymbol{I}$ is the identity matrix, it is readily verified that $\boldsymbol{e}$ has the desired covariance  
$E(\boldsymbol{e}\boldsymbol{e}^T) = \boldsymbol{C}$.
		
In order to make a joint solution the Hipparcos normal equations are reconstructed from the covariance inverse, $\boldsymbol{N}_{HIP} = \boldsymbol{C}_{HIP}^{-1}$.
The right-hand side of the Hipparcos
normal equations needs to be chosen such that solving for the Hipparcos information only, the update
would bring the current parameter values back to the original Hipparcos values.
This can be done by calculating a vector $\boldsymbol{d}$ defined as the difference
between the original Hipparcos source parameters (subscript o) and the current values (subscript c)
and choosing the right-hand side $\boldsymbol{b}_{\textrm{HIP}}$ as
\begin{equation}
\boldsymbol{b}_{\textrm{HIP}} = \boldsymbol{N}_{\textrm{HIP}}\boldsymbol{d} =  \boldsymbol{C}^{-1}_{\textrm{HIP}} \begin{pmatrix}
	(\alpha_o - \alpha_c) \cos\delta_o \\
	\delta_o - \delta_c\\
	\pi_o - \pi_c\\
	\mu_{\alpha \star o} - \mu_{\alpha \star c}\\
	\mu_{\delta o} - \mu_{\delta c}
    \end{pmatrix}.
\end{equation}
Solving $\boldsymbol{N}_\textrm{HIP} \boldsymbol{x} = \boldsymbol{b}_{\textrm{HIP}}$ gives $\boldsymbol{x} = \boldsymbol{d}$ and the application of this update to the current parameters obviously recovers the Hipparcos parameters.
\section{Results} 
Table \ref{michalik_results} shows results of simulation
runs. As expected the combination of Hipparcos and Nano-JASMINE gives a great
improvement in proper motions. Additionally we show that our proposed joint
solution performs significantly better than the conventional catalogue
combination method.  This can be understood as follows. The astrometric
parameters in the Hipparcos (or Nano-JASMINE) catalogue are correlated. The large
improvement of the proper motions therefore brings some improvement
also to the other parameters, provided that the correlations are
properly taken into account. This is the case for the joint solution,
 but not for the conventional combination.

\begin{table}
\begin{tabular}{@{}rrrrrr@{}} 
&\multicolumn{2}{c}{Position @J2015} & \multicolumn{1}{c}{Parallax} & \multicolumn{2}{c}{Proper motions} \\ 
&\multicolumn{2}{c}{[mas]} & \multicolumn{1}{c}{[mas]} & \multicolumn{2}{c}{[mas/year]} \\ 
\multicolumn{1}{r}{\textbf{mag $\boldsymbol{<7.5$, $\sim15\,000}$ stars}}& \multicolumn{1}{c}{$\alpha$} & \multicolumn{1}{c}{$\delta$} & \multicolumn{1}{c}{$\pi$} & \multicolumn{1}{c}{$\mu_{\alpha \star}$} & \multicolumn{1}{c}{$\mu_{\delta}$}\\

\midrule[\heavyrulewidth]
	Hipparcos only (Hip) & 18.19 & 14.84 & 0.80 & 0.77 & 0.63 \\
	Nano-JASMINE only (NJ) & 2.56 & 2.54 & 3.05 & 4.65  &  4.50 \\
	\midrule[0.2pt]
	Conventional combination Hip + NJ & 2.54 & 2.51 &0.77& 0.111 & 0.110 \\
	\textbf{Joint solution Hip + NJ} & \textbf{2.41} & \textbf{2.40} & \textbf{0.75} &\textbf{ 0.108} & \textbf{0.105}\\
	\midrule[0.2pt]
	Improvement of joint solution & 5.2\% & 4.4\% & 3.5\% & 3.2\% &  4.4\% \\
\bottomrule

\multicolumn{6}{c}{~}\\
\multicolumn{1}{r}{\textbf{mag $\boldsymbol{<11.5$, $\sim117\,000}$ stars}}&&&&&\\
\midrule[\heavyrulewidth]
	Hipparcos only (Hip) & 27.06 & 22.35 & 1.18 & 1.14& 0.94\\
	Nano-JASMINE only (NJ) & 4.57 & 4.53 & 5.43 & 8.38 & 8.02 \\
	\midrule[0.2pt]
	Conventional combination Hip + NJ & 4.51 & 4.44 & 1.15 & 0.197 & 0.194\\
	\textbf{Joint solution Hip + NJ} & \textbf{4.43} & \textbf{4.26} & \textbf{1.11} & \textbf{0.188} & \textbf{0.185}\\
	\midrule[0.2pt]
	Improvement of joint solution & 1.8\% & 3.9\% & 4.0\% &  4.5\% & 4.5\% \\
\bottomrule
\end{tabular}
\caption{Conventional catalogue combination vs. a joint astrometric solution, for a subset of bright stars and for all Hipparcos stars.
	Simulations of Nano-JASMINE are based on a conservative observation
performance model and an optimal scanning law. The positions from
Hipparcos have been propagated to the Nano-JASMINE mid-mission epoch J2015.
	\label{michalik_results}}
\end{table}

\subsection{Future work}
We are planning to extend our studies of catalogue combination by simulating
the improvements that can be gained by applying our method to catalogues from
simulated Gaia and Nano-JASMINE data together with the Hipparcos
and the Tycho-2 catalogues.

Furthermore, the goodness-of-fit of the combined solution is sensitive to small
deviations of the stellar motions from the assumed (rectilinear) model.
We are investigating how this can be used to identify binary candidates
with orbital periods of decades to centuries. This will contribute to
the census of the binary population within a few hundred parsecs from
the sun by filling a difficult-to-observe gap between the shorter period
spectroscopic binaries and the longer period visually resolved systems.

\acknowledgements Our research is kindly supported by the Swedish National
Space Board and the European Space Agency.

\bibliography{P100}

\end{document}